\documentclass{article}
\usepackage{spconf,amsmath,graphicx}
\usepackage{setspace}
\usepackage{amssymb}
\usepackage{float}
\usepackage{booktabs}
\usepackage{bibspacing}
\usepackage{color}
\usepackage{multirow}

\title{FSD: An Initial Chinese dataset for Fake Song Detection}
\name{Yuankun Xie, Jingjing Zhou, Xiaolin Lu, Zhenghao Jiang, Yuxin Yang, Haonan Cheng, Long Ye\sthanks{Long Ye is corresponding author. The song demo can be found at https://github.com/xieyuankun/FSD-Dataset.}}
\address{
State Key Laboratory of Media Convergence and Communication, \\
	Communication University of China, Beijing 100024, China\\
}

\begin{document}
%
\maketitle
\begin{abstract}
Singing voice synthesis and singing voice conversion have significantly advanced, revolutionizing musical experiences. However, the rise of “Deepfake Songs” generated by these technologies raises concerns about authenticity. Unlike Audio DeepFake Detection (ADD), the field of song deepfake detection lacks specialized datasets or methods for song authenticity verification.
In this paper, we initially construct a Chinese Fake Song Detection (FSD) dataset to investigate the field of song deepfake detection. The fake songs in the FSD dataset are generated by five state-of-the-art singing voice synthesis and singing voice conversion methods. Our initial experiments on FSD revealed the ineffectiveness of existing speech-trained ADD models for the task of song deepfake detection. Thus, we employ the FSD dataset for the training of ADD models. We subsequently evaluate these models under two scenarios: one with the original songs and another with separated vocal tracks. Experiment results show that song-trained ADD models exhibit a 38.58\% reduction in average equal error rate compared to speech-trained ADD models on the FSD test set.
\end{abstract}
\section{Introduction}
Singing voice synthesis \cite{liu2022diffsinger,chen2020hifisinger,ren2020deepsinger} and singing voice conversion \cite{jayashankar2023self,polyak2020unsupervised,deng2020pitchnet} have witnessed remarkable advancements in recent years, revolutionizing the way we perceive and enjoy music. This advancement, while captivating, raises alarming issues related to authenticity and trustworthiness. The growing capabilities of singing voice synthesis and singing voice conversion enable the creation of fake singing voices, commonly referred as ``Deepfake Songs," which can imitate the style, timbre, and emotion of real singers with astonishing accuracy. This poses a considerable threat to the integrity of audio content, artistic integrity, and the credibility of vocal performances.

In light of these challenges, the necessity for robust and reliable methods of detecting and verifying the authenticity of songs has become increasingly apparent. The field of song deepfake detection has emerged as a crucial area of research. song deepfake detection task aims to develop techniques that can differentiate between genuine and synthetic songs, thereby safeguarding the integrity of musical works, protecting the reputation of artists, and maintaining the trust of listeners. However, to the best of our knowledge, there are currently no dedicated song deepfake detection datasets or methods for detecting the authenticity of songs. 

For a similar area to song deepfake detection, the field of Audio DeepFake Detection (ADD) is relatively comprehensive and boasts an abundance of detection datasets and methods. In order to propel the research in ADD, numerous datasets have been introduced, including those derived from competitions such as the ASVspoof series \cite{delgado2021asvspoof} and ADD challenge series \cite{yi2022add}. In terms of methods, models like AASIST \cite{jung2022aasist} and those based on Wav2Vec2 (W2V2) \cite{baevski2020wav2vec} have achieved equal error rate (EER) below 1\% in single-domain scenarios \cite{eom22_interspeech, martin2022vicomtech}. However, to our knowledge, there has been no investigation of whether these methods can effectively detect the authenticity of songs. 

Judging the authenticity of songs poses a significant challenge. Currently, ADD models can only detect domain-specific datasets in a single domain, lacking robustness and generalizability to out-of-domain datasets \cite{muller22_interspeech}. For singing songs, they are created by mixing vocal tracks and instrumental tracks. Vocals exhibit distinctive attributes, including varying pitch range and modulation, distinguishing them from typical speech which may lead to misjudgment for current ADD methods. On the other hand, the presence of instrumental tracks in songs can be perceived as “interfering noise” by current ADD approaches. It is significant to investigate their effect on song deepfake detection task.

To carry out the relevant research, the construction of the dataset is the first step. In our paper, we introduce a Chinese fake song detection (FSD) dataset. For the fake songs, they totally cover 5 types, which are the popular and expressive generated methods. We extract the instrumental track from real songs and mix it with the fake singing voice to create the final fake song. We also employ state-of-the-art (SOTA) ADD methods to evaluate the FSD dataset under two conditions: the original songs, and the vocal tracks isolated through separation technique. Our main contributions can be summarized as follows:
\begin{itemize}
	\item We initially present FSD dataset, specifically designed for investigating song deepfake detection task.
	\item Leveraging the proposed FSD dataset and audio source separation strategy for training, we achieve a 38.58\% reduction in EER on the FSD test set when compared to speech-trained ADD algorithms.
\end{itemize}

\section{Dataset Design}
The FSD dataset comprises a total of 200 real songs and 450 fake songs. In this section, we will introduce the process of generating the fake songs. 
\subsection{Fake singing voice generation}
We select 5 different representative singing voice synthesis and singing voice conversion methods, denoted as F01 to F05, to generate the fake songs.

\noindent \textbf{F01: SO-VITS\footnote{https://github.com/svc-develop-team/so-vits-svc}.}  This is a project differs fundamentally from VITS \cite{kim2021conditional}, as it focuses on singing voice conversion rather than Text-to-Speech. 
The singing voice conversion model employs the content encoder from SoftVC \cite{van2022comparison} to extract speech features from the source singing voice. These feature vectors are directly input into VITS without necessitating conversion to an intermediary text-based representation. This approach preserves the pitch and intonations of the original singing voice. Additionally, in SO-VITS, the NSF-HiFiGAN vocoder is employed as the vocoder. This modified version of HifiGAN \cite{kong2020hifi}, based on the neural source filter \cite{wang2019neural}, effectively addresses the problem of sound interruptions.

\noindent \textbf{F02: SO-VITS (NSF-HifiGAN with Snake \cite{ziyin2020neural}).} This generation method, modified from F01, introduces modifications to the decoder of VITS, specifically, the vocoder component of the network. The NSF-HiFiGAN vocoder is optimized and integrated, leveraging a novel activation function called "Snake." This activation function is designed to achieve the desired periodic inductive bias, allowing the network to learn periodic functions while retaining the favorable optimization properties associated with ReLU-based activations.
	
\noindent \textbf{F03: SO-VITS (with shallow diffusion \cite{liu2022diffsinger}).} To address the issue of electrical sound problems, this approach incorporates a solution within VITS. Specifically, we trained a separate shallow diffusion model to further enhance the quality of mel-spectrogram. During this training process, we employed a pre-trained NSF-HifiGAN to convert mel-spectrogram into audio.

\noindent \textbf{F04: DiffSinger \cite{liu2022diffsinger}.} This is an acoustic model for singing voice synthesis task based on the diffusion probabilistic model \cite{ho2020denoising}. DiffSinger commences generation at a shallow step determined by the intersection of the diffusion trajectories of the ground-truth mel-spectrogram and the one predicted by a simple mel-spectrogram decoder. Additionally, DiffSinger introduces a boundary prediction technique to dynamically locate and ascertain this intersection point.

\begin{figure}[!t]
	\centering
	\includegraphics[width= 3.2in]{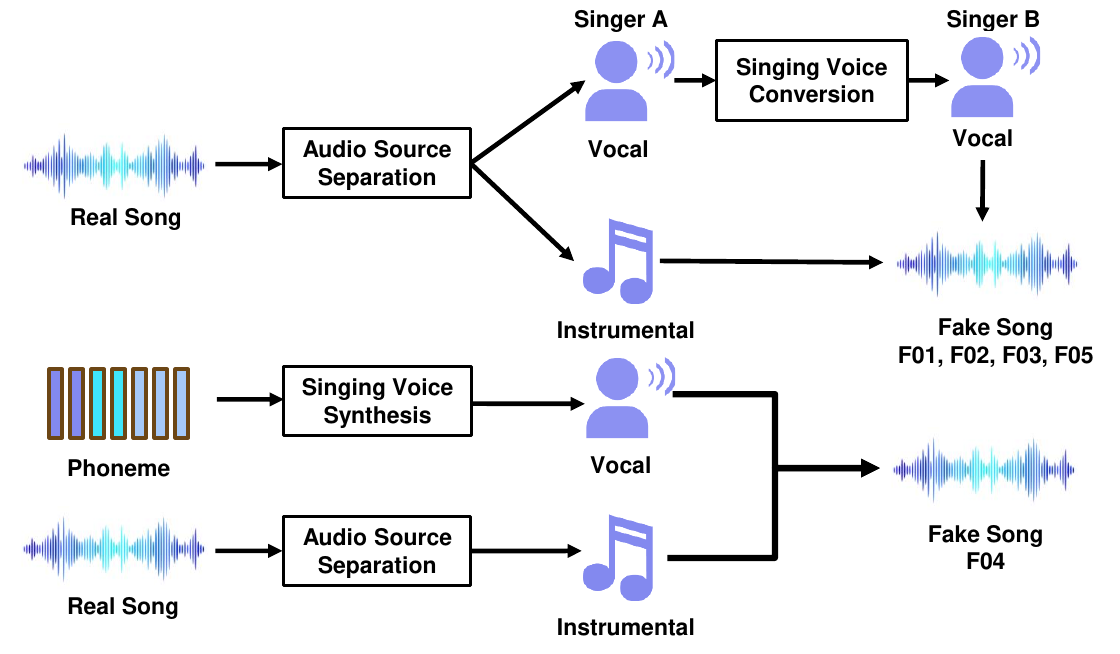}
	\hfil
	\caption{The pipeline of fake song generation.}
	\label{fig:pipeline}
\end{figure}

\noindent \textbf{F05: RVC\footnote{https://github.com/RVC-Project/Retrieval-based-Voice-Conversion-WebUI}.}  This is a popular project modified from VITS \cite{kim2021conditional}, as it focuses on voice conversion and also shows promising result in singing voice conversion task. RVC utilize the content vector \cite{qian2022contentvec} as the input and reduce tone leakage by replacing the source feature with a training-set feature using top1 retrieval. To address the issue of muted sound, RVC incorporates a powerful high-pitch voice extraction algorithm \cite{wei2023rmvpe}, showing promising result with enhanced efficiency.

\subsection{Fake song generation}
In practice, fake songs not only take the form of fake singing voice but also involve mixing with instrumental track, making detection more challenging. Therefore, in this step, we perform the mixing with the corresponding instrumental track, as illustrated in Fig. \ref{fig:pipeline}. For the F01, F02, F03, and F05 generation types, we initially employ an audio source separation tool\footnote{https://github.com/Anjok07/ultimatevocalremovergui} to isolate the raw audio into vocal and instrumental track. Subsequently, the singing voice conversion method is applied to convert the vocal of singer A to that of singer B. Finally, the generated fake song is produced by combining the vocal track and instrumental track while considering their corresponding amplitudes. For the F04 generation type, we use Pypinyin\footnote{https://github.com/mozillazg/python-pinyin} to convert Chinese lyrics from real songs to phonemes and let professional annotator to correct the automatically converted phonemes to standard phonemes by listening to the audio. Then, singing voice synthesis method is used to generate the fake singing voice from the phoneme information. Lastly, we mix the separated instrumental track with the fake singing voice to generate the fake song.
\begin{table*}[t]
	\caption{Details for fake singing voice generation in FSD dataset.}
	\centering
	\renewcommand{\arraystretch}{0.2}
	\setlength{\tabcolsep}{5pt}
	\begin{tabular}{cccccccc}
		\toprule
		Method &Type&Synthesizer &Vocoder&Source&Target&Songs &Hours\\
		\midrule
		F01  &SVC&SO-VITS &NSF-HifiGAN&M1/F1 &bass-1/opencpop&100&5.88\\
		\midrule
		F02  &SVC&SO-VITS &NSF-Snake-HifiGAN&M1/F1 &bass-1/opencpop&100&5.88\\
		\midrule
		F03  &SVC&SO-VITS-Diff &NSF-HifiGAN &M1/F1 &bass-1/opencpop&100&5.88\\
		\midrule
		F04  &SVS&DiffSinger &NSF-HifiGAN &Phoneme &M2/F2&100&5.07\\
		\midrule
		F05  &SVC&RVC &NSF-HifiGAN &M3/F3 &tenor-7/alto-5 &50&3.55\\
		\bottomrule
	\end{tabular}
	\label{tab:details}
\end{table*}
\begin{figure*}[!t]
	\centering
	\includegraphics[width= 6 in]{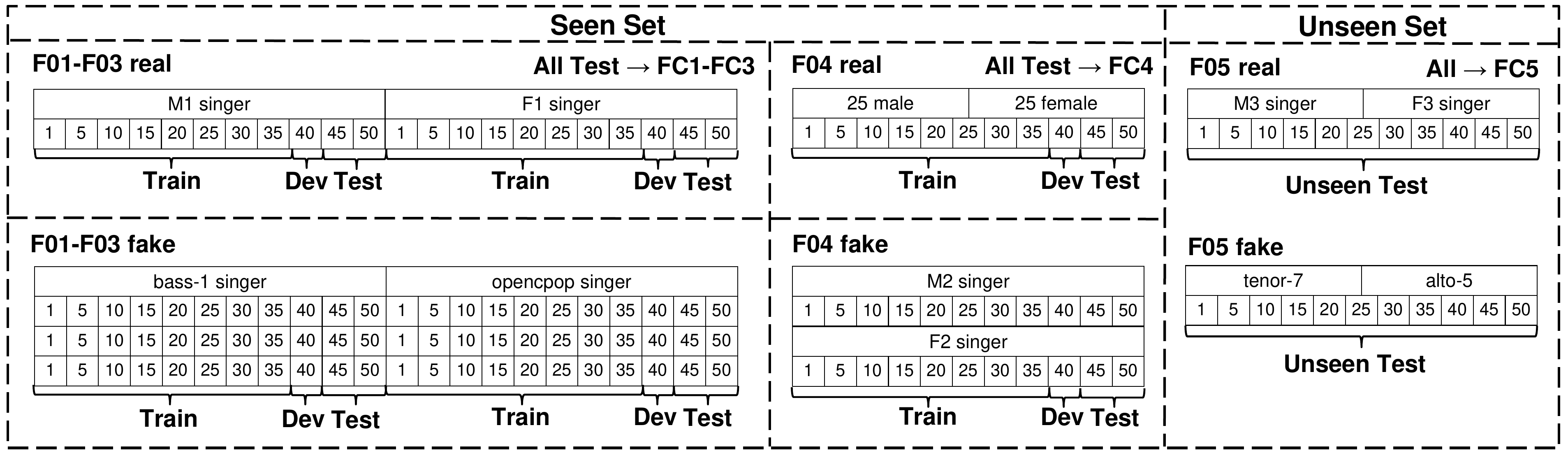}
	\hfil
	\caption{Partition and construction of FSD dataset. The numbers in the table correspond to the indices of the songs.}
	\label{fig:partition}
\end{figure*}
\section{Experiments}
\label{sec:typestyle}
\subsection{Data construction details}
The details for fake singing voice generation are shown in Table \ref{tab:details}. For F01-F03 conditions, we collect 100 real songs including one male (M1) singer and one female (F1) singer as source domain. Subsequently, we employ the bass-1 male singer from the M4singer dataset \cite{zhang2022m4singer} and a female singer from the opencpop dataset \cite{wang2022opencpop} to train the SO-VITS model. This enables the transformation of singer M1 into bass-1 and singer F1 into opencpop. For F04 conditions, we collect 50 real songs including 25 different male singers, and 25 female singers and extract the phoneme information. Then, for training, we utilize M4singer dataset to train a basic DiffSinger model and finetune on one male singer (M2) and one female (F2) singer. For F05 conditions, we collect 50 real songs which include one male (M3) singer and one female (F3) singer as source domain. Then, we train the RVC model using the tenor-7 male singer and alto-5 female singer, allowing the conversion from singer M3 to tenor-7 and singer F3 to alto-5. 

\subsection{Experiments settings}
To assess the performance of current ADD models on the FSD dataset, we segmented the fake songs into 4-second segments. We further divided the dataset into training, development, and evaluation sets as shown in Fig. \ref{fig:partition}. The division is based on the sequential numbering of the source domain songs, with the first 35 songs allocated to the training set, songs 36 to 40 to the development set, and songs 41 to 50 to the evaluation set. We create five testing conditions, denoted as FC1 to FC5, corresponding to different fake generation types. Each testing condition includes both the synthetic fake song and the genuine song from their source domain. For example, FC1 comprises the 41th to 50th songs generated by F01 method, as well as the 41th to 50th songs from the source domain, which are authentic. It is worth noting that we employed FC5 as an unseen test set to evaluate the generalizability of the model. This implies that there are no FC5 songs in either the training or development sets. The number of segments in each subset is shown in Table \ref{tab:number}.
\begin{table}[t]
	\caption{Number of segments in each subset}
	\centering
	\renewcommand{\arraystretch}{1}
	\setlength{\tabcolsep}{4pt}
	\begin{tabular}{c|c|c|ccccc}
		\hline
		Type &Train &Dev &FC1&FC2&FC3&FC4 &FC5\\
		\hline
		Real  &5261 &795 &1131&1131 &1131&482&3226\\
		\hline
		Fake  &14242 &2094 &1131&1131 &1131&964&3226\\
		\hline
		Total  &19503 &2889 &2262 &2262 &2262&1446 &6452\\
		\hline
	\end{tabular}
	\label{tab:number}
\end{table}
\begin{table*}[t]
	\caption{EER (\%) results on the full FSD test set. “AVG” represents the average EER across FC1-FC5, while “AVG↓” denotes the EER reduction from the speech-trained ADD model to the song-trained ADD model.}
	\centering
	\renewcommand{\arraystretch}{1}
	\setlength{\tabcolsep}{1.5pt}
	\begin{tabular}{c|c|cccccc|c|cccccc|cc}
		\hline
		\multirow{2}{*}{Model} & \multicolumn{7}{c|}{speech-trained ADD model (I)} & \multicolumn{7}{c}{song-trained ADD model (II)} &\multicolumn{1}{|c}{\multirow{2}{*}{AVG↓}}\\
\cline{2-15}
		&19LA &FC1&FC2&FC3&FC4 &FC5&AVG &Dev &FC1&FC2&FC3&FC4 &FC5&AVG \\
		\hline
		AASIST & 0.83	&46.50	&\bf {45.53}	&48.62	&\bf {46.21}	&\bf {41.75}	&\bf {45.72} &\bf8.68	&\bf8.22	&\bf8.66	&\bf8.93	&\bf 16.85	&23.68	&\bf13.27 &\bf32.45
		\\
		\hline
		Mel-LCNN  &2.68	&47.92	&47.74	&49.43	&47.30	&44.82	&47.44 &10.89	&11.40	&11.31	&12.82	&21.57	&20.42	&15.50&31.94
		\\
		\hline
		W2V2-LCNN  &\bf {0.69}	&\bf {43.23}	&46.77	&\bf {48.01}	&49.01	&47.76	&46.96 &13.65	 &16.09	 &16.89	 &16.90	 &18.46	 &\bf {23.24}	 &18.32&28.64
		\\
		\hline
	\end{tabular}
	\label{tab:fullset}
\end{table*}
\begin{table*}[t]
	\caption{EER (\%) results on the separated vocal track of FSD test set. “AVG” represents the average EER across FC1-FC5, while “AVG↓” denotes the EER reduction from the speech-trained ADD model to the song-trained ADD model.}
	\centering
	\renewcommand{\arraystretch}{1}
	\setlength{\tabcolsep}{1.5pt}
	\begin{tabular}{c|c|cccccc|c|cccccc|cc}
		\hline
		
		\multirow{2}{*}{Model} & \multicolumn{7}{c|}{speech-trained ADD model (I)} & \multicolumn{7}{c}{song-trained ADD model (II)} &\multicolumn{1}{|c}{\multirow{2}{*}{AVG↓}}\\
        \cline{2-15}
		&19LA &FC1&FC2&FC3&FC4 &FC5&AVG &Dev &FC1&FC2&FC3&FC4 &FC5&AVG& \\
		\hline
		AASIST  &0.83 &51.45 &50.04	 &52.87	 &48.96	 &\bf36.36	 &\bf47.94& 7.65	&7.07	&7.26	&\bf6.81	&\bf12.24	&31.25	&12.93 &35.01
		\\
		\hline
		Mel-LCNN  &2.68	&48.80	&\bf48.27	&49.42	&\bf48.13	&46.47	&48.22 &14.73	 &18.30	 &18.56	 &17.77	 &25.31	 &29.82	 &21.95 &26.27
		\\
		\hline
		W2V2-LCNN  &\bf0.69	&\bf47.39	&49.60	&\bf48.01	&49.58	&45.93	&48.10 &\bf6.79&	\bf6.98&	\bf7.25	&7.07	&12.44	&\bf13.85	&\bf9.52 &\bf38.58
		\\
		\hline
	\end{tabular}
	\label{tab:vocalset}
\end{table*}
\subsection{Implementation details}
We comprehensively evaluate the FSD dataset using SOTA ADD methods, namely AASIST \cite{jung2022aasist} and LCNN \cite{lavrentyeva2019stc}. In terms of features, we utilized three distinct types: raw audio, mel spectrogram, and W2V2 representations. For the mel spectrogram, we extracted a 128-dimensional mel spectrogram. For W2V2, we employed the Wav2Vec-XLS-R\footnote{https://huggingface.co/facebook/wav2vec2-xls-r-300m} model with frozen parameters, extracting the 1024-dimensional last hidden states as the feature representation. All models are trained for 100 epochs. We used Adam optimizer with a learning rate of $10^{-4}$ and cosine annealing learning rate decay.

\section{Results and Discussion}
\subsection{Speech-trained ADD models results}
We first assessed the efficacy of the present speech-trained ADD models using the FSD dataset, as outlined in Table \ref{tab:fullset}(I). Concretely, we conducted training on three distinct models: AASIST, Mel-LCNN, and W2V2-LCNN, utilizing the ASVspoof2019LA (19LA) \cite{todisco19_interspeech} training set. From the results, we can observe that they achieve promising result in 19LA test sets and W2V2-LCNN achieve the lowest EER with 0.69\%. Then, for FSD test sets, the results are quite unfavorable, with all EER falling between 40\% and 50\%. We suppose that this might be due to the influence of mixing the instrumental tracks. Thus, we conduct a second experiment, where we tested the extracted vocal track of FSD test set through audio source separation methods. The results are displayed in Table \ref{tab:vocalset}(I). We observe that the results of detecting the separated vocal track are not satisfactory. This could be attributed to the inherent limitations of speech-driven models for cross-domain detection, especially when detecting vocal, which contains richer pitch and rhythm information. Furthermore, both genuine and fake songs underwent vocal track separation using the same method, which involves upsampling convolution. This process might introduce artifacts that affect both real and fake vocals, thereby impacting the final decision. 
\vspace{-0.3cm}
\subsection{Song-trained ADD models results}
Due to the poor performance of speech-trained ADD models on the FSD test set, we proceed to train the baseline models using the FSD training set. The results are presented in the Table \ref{tab:fullset}(II). In comparison to the speech-trained models, the AASIST, Mel-LCNN, and W2V2-LCNN models exhibited reductions in average EER of 32.45\%, 31.94\%, and 28.64\%, respectively, on the FSD full test set. Among these, AASIST achieved the best average EER of 13.27\%. Since the validation set includes domain information for FC1-FC4, their EER performances are similar. However, the genuine source domain of FC4 includes real songs from out-of-domains, resulting in slightly worse performance than FC1-FC3. For the unseen condition FC5,  the performance of three baseline models is not particularly strong, which indicate that the current models lack the generalization ability to detect out-of-domain songs effectively.

Similarly, we also evaluate the performance of the separated vocal tracks as shown in Table \ref{tab:vocalset}(II). 
In comparison with the results for full test set, we observed a significant improvement in performance for the W2V2-LCNN, achieving a lowest EER of 9.52\% and a highest 38.58\%  average EER reduction compared to the speech-trained W2V2-LCNN. This indicates that W2V2 features hold certain advantages in vocal-level discrimination similar to their performance in speech deepfake detection. Furthermore, under the unseen condition of FC5, W2V-LCNN maintains a low EER value of 13.85\%, which demonstrates the strong generalization characteristics of W2V2 features even when applied to vocal domain.

\vspace{-0.2cm}
\section{Conclusion}
\vspace{-0.2cm}
In this paper, we propose fake song detection (FSD) dataset for song deepfake detection task. We constructed the novel song dataset utilizing five of the most popular and expressive generation methods. With this dataset, we first investigate the performance of SOTA ADD methods for song deepfake detection task. Due to the unsatisfactory outcomes achieved by speech-trained ADD methods, we turned to training with the FSD dataset. Specifically, we employed both the original songs and the vocal tracks separated through audio source separation for training and testing. Experimental results reveal the effectiveness of training with the separated vocal track, with the W2V2-LCNN model achieving the lowest EER of 9.52\%. Future work will encompass the incorporation of a wider range of generation methods and the development of novel detection algorithms specifically to the task of song deepfake detection.

\small
\bibliographystyle{IEEEbib}
\bibliography{strings,refs}

\end{document}